\documentclass[a4paper]{jpconf}
\usepackage{graphicx}

\newcommand\al{\alpha}
\newcommand\ga{\gamma}
\newcommand\de{\delta}
\newcommand\ep{\epsilon}
\newcommand\et{\eta}
\newcommand\la{\lambda}
\newcommand\ph{\phi}
\newcommand\vp{\varphi}
\newcommand\ps{\psi}
\newcommand\bx{{\bf x}}
\newcommand\pa{\partial}
\newcommand\beq{\begin{equation}}
\newcommand\eeq{\end{equation}}
\newcommand\X{\times}
\newcommand\fr{\frac}
\newcommand\cL{\mathcal{L}}
\newcommand\cV{\mathcal{V}}
\newcommand\rms[1]{_{\mathrm{#1}}}
\newcommand\<{\langle}
\renewcommand\>{\rangle}

\begin{document}
\title{History of electroweak symmetry breaking}

\author{TWB Kibble}

\address{Blackett Laboratory, Imperial College, London SW7 2AZ}

\ead{t.kibble@imperial.ac.uk}

\begin{abstract}
In this talk, I recall the history of the development of the unified electroweak theory, incorporating the symmetry-breaking Higgs mechanism, as I saw it from my standpoint as a member of Abdus Salam's group at Imperial College.  I start by describing the state of physics in the years after the Second World War, explain how the goal of a unified gauge theory of weak and electromagnetic interactions emerged, the obstacles encountered, in particular the Goldstone theorem, and how they were overcome, followed by a brief account of more recent history, culminating in the historic discovery of the Higgs boson in 2012.
\end{abstract}

\section{Introduction}

I have always believed that it was a stroke of great good fortune that I was able to join the theoretical physics group at Imperial College less than three years after it was founded by Abdus Salam.  Salam already had a top-rank international reputation for his work on renormalization theory.  In 1959, the year I joined the group, he was elected as the youngest Fellow of the Royal Society, at the age of 33.  He was a man of immense vivacity and charisma, and a most inspiring leader of the group.

It was a wonderful place for a young postdoctoral fellow setting out on a research career.  Imperial was a very lively place, with numerous visitors from around the world, including Murray Gell-Mann, Stanley Mandelstam, Steven Weinberg, Ken Johnson and many others.

\section{Physics in the 1950s}

So what were we doing?  The great success of the immediate post-war period had been the development of renormalization theory.  This banished the infinities that had plagued quantum electrodynamics (QED), allowing calculations beyond the lowest order of approximation.  With experiments measuring the magnetic moment of the electron and the Lamb shift in hydrogen, QED rapidly became the most accurately verified theory in the history of physics.

Following this triumph, the natural next goal was to find similarly successful theories of the other fundamental interactions, the strong nuclear force that binds the constituents in atomic nuclei and the weak nuclear interactions responsible for radioactive beta decay, which also play a key role in the energy generation mechanism in the Sun.  Even better, one might hope for a unified theory of all of these; that is a goal that is still ahead of us.  We hoped too at some stage to incorporate the fourth fundamental force, gravity, but that of course is so weak as to be negligible on particle-physics scales.  That too is still a work in progress.

Most interest initially fell on the strong interactions, and there were candidate quantum field theories that were widely discussed, in particular Yukawa's meson theory, in which the $\pi$-mesons or pions played the role of the force carriers.  The problem there, however, was that it was hard to do any meaningful calculations, because the only known calculational technique, perturbation theory, does not work if the `small' dimensionless parameter, the analogue of the fine structure constant of QED, is of order unity, as appeared to be the case.

As a result, quantum field theory fell out of favour.  Many physicists abandoned it in favour of S-matrix theory, based on a study of the complex analytic properties of scattering amplitudes, working on concepts like Regge poles and dispersion relations.  There were, however, places, like Imperial College, where the flag of field theory was kept flying --- another was Harvard.  Certainly Salam never gave up his belief that a unified field theory of all the interactions would eventually be found.

However, the difficulties of calculating with strong interactions did lead some people to speculate that weak interactions might be a more promising initial target.

\section{Approximate symmetries}

The experimenters had been busy too.  Using cloud chambers and later bubble chambers, they discovered a huge array of new `elementary' particles.  Over a few years the number went from a handful to a hundred or more.

To try to make sense of this zoo of particles, theorists looked for approximate symmetries, trying to find patterns among the particles, much as chemists had done in developing the periodic table in the nineteenth century.

The first of these symmetries was \emph{isospin}, originally proposed by Heisenberg \cite{Heisenberg:1932}.  He noted that protons and neutrons are in many ways very similar; they have the same spin, the same strong interactions and nearly the same mass.  The name `isospin' does not imply any physical relation to spin angular momentum, but stems from a mathematical similarity with the description of electron spin.  Heisenberg  suggested that the proton and neutron might be regarded as two different states of a single entity, the \emph{nucleon}, and that there is a symmetry under which one can rotate the proton into a neutron or a combination of the two, much as one can rotate the up-spin state of an electron into the down-spin state or something in between.  In fact isospin and spin are both described by SU(2) groups.

This is of course an \emph{approximate} symmetry: the proton and the neutron are not identical;.  The proton has an electric charge, the neutron does not.  The symmetry is \emph{broken} by the electromagnetic interaction.  However, on a nuclear scale, the electromagnetic interaction is relatively weak compared to the strong nuclear force, so isospin is still a fairly good approximation.  It proved very useful for example in classifying low-lying energy levels of light nuclei.

Later, Kemmer \cite{Kemmer:1938} showed that the symmetry could be extended to the other known strongly-interacting particles, the pions, which form an isospin triplet of charge states $(\pi^+,\pi^0,\pi^-)$, analogous to the spin states of a particle of spin $S=1$ (in units of $\hbar$), and that one could write down a version of Yukawa's theory that was invariant under the isospin transformations.  The isospin symmetry is now seen as fundamentally a symmetry between the two lightest \emph{quarks}, the \emph{up} and \emph{down} quarks ($u$ and $d$), the constituents of nucleons and pions.

The next step was taken in 1961, independently by Gell-Mann \cite{Gell-Mann:1961} and by Yuval Ne'eman \cite{Ne'eman:1961}, who was at the time a student of Salam's at Imperial College.  They proposed a larger, more approximate, SU(3) symmetry, and showed that the newly discovered \emph{hadrons}, or strongly interacting particles, could be grouped into two-dimensional patters of octets and decuplets, forming multiplets related by SU(3) symmetry transformations.  Gell-Mann called this scheme the \emph{eightfold way}.  It is now seen as a symmetry of the \emph{three} lightest quarks, up, down and \emph{strange} ($u$, $d$, $s$).

\section{Gauge theories}

Quantum electrodynamics is a \emph{gauge theory}.  This means it has a special kind of symmetry, a \emph{local} symmetry.  In the quantum mechanics of a single electron, the physical situation is unchanged if the electron wave function is multiplied by a phase factor; we have invariance under the transformation
 \beq \ps(x)\to\ps(x)e^{i\al} \eeq
for any constant $\al$.  But in QED we have a more general invariance for the electron field, in which $\al$ becomes a function of space and time, \emph{provided} that a corresponding transformation is applied to the electromagnetic potentials $A_\mu(x)$; specifically
 \beq \ps(x)\to\ps(x)e^{ie\al(x)},\qquad A_\mu(x)\to A_\mu(x)-\pa_\mu\al(x). \eeq
This is a vitally important feature: the success of the renormalization procedure relies on its existence.

For this reason, many people, including Salam, believed that viable theories of the other interactions should also be gauge theories.  The first gauge theory beyond QED was the Yang-Mills theory \cite{Yang:1954ek}, proposed in 1954, a gauge theory based on the isospin SU(2) symmetry.  The same theory was in fact written down in the same year by a student of Salam's, Ronald Shaw, though he never published it except as a Cambridge University PhD thesis \cite{Shaw:1955}.  Ultimately this did not prove to be the correct theory of strong interactions --- that is a very different kind of gauge theory, \emph{quantum chromodynamics} (QCD) --- but it is nevertheless the foundation for all later work on gauge theories.

Because isospin is an approximate symmetry, it must be broken in some way.  But that immediately led to problems, because simply adding symmetry-breaking terms destroys many of the nice properties of gauge theories.  So people began to ask the question: could this be some sort of \emph{spontaneous} symmetry breaking, avoiding the need to add explicit symmetry-breaking terms?  This is a subject I shall return to shortly.

\section{The goal of unification}

Because of the difficulties already mentioned of calculating with a strong-interaction theory, people began to think that perhaps the weak interactions would be a better immediate target, especially after the development in 1958 of the $V-A$ theory by Feynman and Gell-Mann \cite{Feynman:1958} and by Sudarshan and Marshak \cite{Sudarshan:1958vf}, which showed that they could be seen as proceeding via the exchange of spin-1 $W^\pm$ bosons, just as electromagnetic interactions proceed via the exchange of spin-1 photons.  This led Schwinger \cite{Schwinger:1957} to suggest a gauge theory of weak interactions mediated by $W^+$ and $W-$ exchange.  He even asked: could there be a unified theory of weak and electromagnetic interactions, involving three gauge bosons, $W^+$, $W^-$ and the photon $\ga$?

But this idea immediately ran into difficulty.  If there is in fact a symmetry between these three gauge bosons, it clearly must be severely broken, because there are major differences between them.  The fact that the weak interactions are short-range implies that the bosons that carry this interaction must be very massive, whereas the photon, mediating the long-range electromagnetic force, is massless.  It was clear that if the interaction strengths are comparable, the $W^\pm$ must have masses of the order of 100 GeV.

There was another key difference: it was known that the weak interactions do not conserve parity --- they violate mirror symmetry --- whereas the electromagnetic interactions are parity-conserving.  So how could there be a symmetry between the two?

This latter problem was solved in 1961 by Glashow \cite{Glashow:1961tr}, who proposed an extended model with a larger symmetry group, SU(2)$\X$U(1), and a fourth gauge boson $Z^0$.  He showed that by an intriguing mixing mechanism between the two neutral gauge bosons, one could end up with one boson ($\ga$) with parity-conserving interactions and three that violate parity, $W^+,W^-$ and $Z^0$.

In 1964, Salam and his long-term collaborator John Ward, apparently unaware of Glashow's work, proposed a very similar model also based on SU(2)$\X$U(1) \cite{Salam:1964}.

But in all these models, the symmetry breaking, giving the $W$ and $Z$ bosons masses, had to be inserted by hand, and theories of spin-1 bosons with explicit masses were well known to be non-renormalizable and thus unphysical.  The big question therefore was: could this be a case of spontaneous symmetry breaking?

\section{Spontaneous symmetry breaking}

Spontaneous symmetry breaking occurs when the ground state or vacuum, or equilibrium state of a system does not share the underlying symmetries of the theory.  It is ubiquitous in condensed matter physics, associated with phase transitions.  Often, there is a high-temperature symmetric phase and a critical temperature below which the symmetry breaks spontaneously.  A simple example is crystalization.  If we place a round bowl of water on a table, it looks the same from every direction, but when it freezes the ice crystals form in specific orientations, breaking the full rotational symmetry.  The breaking is spontaneous in the sense that, unless we have extra information, we cannot predict in which directions the crystals will line up.  A similar thing happens when a ferromagnet is cooled through its Curie point; it may acquire a magnetization in one direction or the reverse, but we cannot say in advance which it will choose.

In 1960, Nambu \cite{Nambu:1960tm} pointed out that gauge symmetry is broken in a superconductor when it goes through the transition from normal to superconducting, and that this gives a mass to the plasmon, although this view was still quite controversial in the superconductivity community (see also Anderson \cite{Anderson:1963}).  Nambu suggested that a similar mechanism might give masses to elementary particles.  The next year, with Jona-Lasinio \cite{Nambu:1961tp}, he proposed a specific model, though not a gauge theory, a four-fermion interaction based on the interaction Lagrangian
 \beq \cL\rms{int} = g[(\bar\ps\ps)^2-(\bar\ps\ga_5\ps)^2]. \eeq
Here the phase symmetry $\ps\to e^{i\al}\ps$ is unbroken, but the chiral symmetry $\ps\to e^{\al\ga_5}\ps$ is spontaneously broken because $\<\bar\ps\ps\>\ne0$.  This means that there is a non-zero mass $m_\ps$.

The model has a significant feature, a massless pseudoscalar particle, which Nambu and Jona-Lasinio tentatively identified with the pion.  To account for the non-zero (though small) pion mass, they suggested that the chiral symmetry was not quite exact even before the spontaneous symmetry breaking.

Attempts to apply this idea to symmetry breaking of fundamental gauge theories however ran into a severe obstacle, the \emph{Goldstone theorem}.

\section{Nambu--Goldstone bosons}

As in the Nambu--Jona-Lasinio model, the spontaneous breaking of a continuous symmetry often leads to the appearance of massless spin-0 particles.  The simplest model that illustrates this is the \emph{Goldstone model} \cite{Goldstone:1961eq}, a model of a complex scalar field with the Lagrangian density
 \beq \cL = \pa_\mu\ph^*\pa^\mu\ph - \cV, \eeq
where $\cV$ is the \emph{sombrero potential},
 \beq \cV = \fr{1}{2}\la\left(\ph^*\ph-\fr{1}{2}\et^2\right)^2,\label{V} \eeq
where $\la$ and $\et$ are positive constants.  This Lagrangian is clearly invariant under global phase rotations, $\ph(x)\to\ph(x)e^{i\al}$.

The key feature of the potential is that it has a \emph{maximum} at $\ph=0$.  Consequently, in the vacuum state, the expectation value of $\ph$ is not zero, but lies somewhere on the circle of minima:
 \beq \<0|\ph(x)|0\> = \fr{1}{\surd 2}\et e^{i\al}. \label{vev}\eeq
This breaks the symmetry.  We have a continuous infinity of degenerate vacuum states, distinguished by the value of $\al$, though the underlying symmetry is exact; all are in fact physically equivalent.

Suppose for example we choose $\al$ to be real, and set
 \beq \ph(x) = \fr{1}{\surd 2}[\et + \vp_1(x) + i\vp_2(x)], \label{phi}\eeq
where $\vp_{\{1,2\}}$ are real.  Then, expanding (\ref{V}), we find
 \beq \cV = \fr{1}{2}\la\et^2\vp_1^2 + \mathrm{cubic\ and\ quartic\ terms}. \eeq
Thus $\vp_1$ has a mass but $\vp_2$ is massless.  Looking at the figure, it is easy to see why this is.  The $\vp_1$ mass is determined by the curvature of $\cV$ in the radial direction, but the $\vp_2$ mass vanishes because in the transverse direction $\cV$ is flat.  The $\vp_2$ excitations correspond to space-time dependent variations in the angle $\al$ defining the direction of symmetry breaking.

The appearance of these massless spin-zero \emph{Nambu--Goldstone bosons} was believed to be an inevitable consequence of spontaneous symmetry breaking in a relativistic theory; this is the content of the \emph{Goldstone theorem}.  That is a problem because such massless particles, if they had any reasonable interaction strength, should have been easy to see, but none had been seen.

\section{The Goldstone theorem}

This problem was obviously of great concern to all those who were trying to build a viable gauge theory of weak interactions.  When Steven Weinberg came to spend a sabbatical at Imperial College in 1961, he and Salam spent a great deal of time discussing the obstacles.  They developed a proof of the Goldstone theorem, published jointly with Goldstone \cite{Goldstone:1962es}.

The argument is straightforward.  We start with two assumptions.  First, the symmetry in question corresponds to a current $j^\mu(x)$ which is conserved, i.e,
 \beq \pa_\mu j^\mu=0, \label{cont}\eeq
and which generates the transformations in the sense that the variation of any field $\ph(x)$ is given by
 \beq \de\ph(0) = i\ep\int d^3\bx\,[\ph(0),j^0(0,\bx)], \label{broken}\eeq
where $\ep$ is an infinitesimal parameter.  Second, there is a scalar field $\ph$  whose vacuum expectation value is not invariant under this transformation:
 \beq \<0|\de\ph(0)|0\>\ne 0. \eeq
Now the continuity equation (\ref{cont}) would seem to imply the existence of a conserved charge operator:
 \beq \fr{dQ}{dt}=0,\qquad Q(t)=\int d^3\bx\,j^0(t,\bx). \label{consQ}\eeq
Then the broken symmetry condition (\ref{broken}) can be written
 \beq i\<0|[\ph(0),Q]|0\>=\et\ne 0. \eeq
But if $Q$ is time-independent, then the only intermediate states that can contribute in the commutator are \emph{zero-energy} states, which can only appear if there are massless particles in the theory.

This argument seemed conclusive, with no escape.  Spontaneous symmetry breaking implied massless spin-zero bosons, which should have been easy to see but had not been seen.  On the other hand adding explicit symmetry-breaking terms led to non-remnormalizable theories predicting infinite results.  Weinberg commented `Nothing will come of nothing; speak again', a quotation from \emph{King Lear}.
Fortunately, however, our community was able to speak again.

\section{Resolution of the impasse}

In 1964 a young American postdoctoral fellow arrived at Imperial College, Gerald Guralnik.  He had been a student of Walter Gilbert at Harvard, while Gilbert had earlier been a student of Salam in Cambridge.  Guralnik had been studying the problem of giving masses to gauge bosons, and had already published some ideas about it \cite{Guralnik:1964a,Guralnik:1964b}.  We began collaborating, with another American visitor, Carl Richard Hagen, to find a way round the obstacle posed by the Goldstone theorem.  We, and of course others, eventually succeeded. 

The argument fails in the case of a gauge theory, for quite subtle reasons to which I will return.  The proof is valid, but there is a hidden assumption which, though seemingly natural, is violated by gauge theories.   This was discovered independently by three groups, first Englert and Brout from Brussels \cite{Englert:1964et}, then Higgs from Edinburgh \cite{Higgs:1964ia,Higgs:1964pj} and finally Guralnik, Hagen and myself from Imperial College \cite{Guralnik:1964eu}.  All three groups published papers in Physical Review Letters during the summer and autumn of 1964, papers that were selected by the journal, during its 50th anniversary celebrations as among the most significant of that year.  (Higgs also wrote a paper for Physics Letters.)

The simplest model that illustrates this mechanism is the \emph{Abelian Higgs model}, which is just a gauged version of the Goldstone model, with the Lagrangian
 \beq \cL = D_\mu\ph^*D^\mu\ph - \fr{1}{4}F_{\mu\nu}F^{\mu\nu} - \cV, 
 \label{L} \eeq
where
 \beq D_\mu\ph = \pa_\mu\ph+ieA_\mu\ph, \qquad 
 F_{\mu\nu}=\pa_\mu A_\nu - \pa_\nu A_\mu, \eeq
and $\cV$ as before is the sombrero potential (\ref{V}).  Once again, $\ph$ acquires a non-zero vacuum expectation value (vev), lying somewhere on the circle of minima (\ref{vev}).

Let us again choose $\al=0$, and write $\ph$ in terms of vev plus real and imaginary parts.  It is convenient also to define a new vector field
 \beq B_\mu = A_\mu - \fr{1}{e\et}\pa_\mu\vp_2. \label{B}\eeq
Note that because this has the form of a gauge transformation, we have
 \beq F_{\mu\nu}=\pa_\mu B_\nu - \pa_\nu B_\mu. \eeq
Substituting into (\ref{L}), we find
 \beq \cL = \fr{1}{2}\pa_\mu\vp_1\pa^\mu\vp_1 - \fr{1}{4}F_{\mu\nu}F^{\mu\nu}
 -\fr{1}{2}\la\et^2\vp_1^2 + \fr{1}{2}e^2\et^2 B_\mu B^\mu
 + {\rm cubic\ terms} + \dots, \eeq
Thus we have arrived at a model with no massless particles, only a massive scalar $\vp_1$ and a massive vector, $B_\mu$.  Seemingly almost by magic, the massless gauge and Goldstone bosons have combined to give a massive gauge boson.

But there is more to the story than this.  We have to ask about the original fields $\ph$ and $A_\mu$,  The field equations, in particular the Maxwell equation
 \beq \pa_\mu F^{\mu\nu} = j^\nu = -e^2\et^2 B^\nu + \dots \eeq
are satisfied by any arbitrary value of $\vp_2$ so long as $B_\mu$, given by (\ref{B}), vanishes, or is some other solution.  This is simply a statement of the gauge invariance of the original model.

To tie down not only $B_\mu$ but also the original fields $A_\mu$ and $\vp_2$ we need to impose a gauge condition.  There are two obvious choices.  We could select the Coulomb gauge condition $\pa_k A^k = 0$.  Then $B_\mu=0$ requires that $\vp_2$ also vanishes, or at most is a constant.

Alternatively, we may choose the Lorentz gauge condition $\pa_\mu A^\mu = 0$.  In that case $B_\mu=0$ only requires that $\vp_2$ satisfy the wave equation $\pa_\mu\pa^\mu\vp_2=0$.  In this manifestly covariant gauge, the Goldstone theorem does apply; $\vp_2$ is a massless scalar field.  But the Goldstone boson is a pure gauge mode; $\vp_2$ has vanishing matrix elements between physical states.

\section{Avoiding the Goldstone theroem}

How then did we manage to avoid the Goldstone theorem?  Where did the argument go wrong?

In the proof, I said that the continuity equation (\ref{cont}) implied the existence of a conserved charge, (\ref{consQ}).  But this is only true if we can drop a surface integral at infinity:
 \beq \fr{dQ(t)}{dt} = \int d^3\bx\, \pa_0 j^0(t,\bx) = - \int d^3\bx\, \pa_k j^k(t,\bx)
 = -\int dS_k\, j^k(t,\bx). \eeq
Dropping this is legitimate in a manifestly Lorentz-invariant theory (e.g.\ Lorentz-gauge QED), because what we are really interested in are commutators, and for such a theory the commutators vanish outside the light cone.  But it is \emph{not} legitimate in Coulomb-gauge QED where commutators fall off quite slowly with distance.

In fact, as shown in our paper \cite{Guralnik:1964eu}, it turns out that when the symmetry is spontaneously broken, the integral defined in (\ref{consQ}) is not merely time-varying, but actually \emph{does not exist} as a self-adjoint operator.  For example, in the Abelian Higgs model, it has the form
 \beq Q = -e^2\et^2 \int d^3\bx\,B^0(t,\bx) + \dots \eeq
and this integral is clearly divergent.

Related to the non-existence of $Q$ is the fact that no unitary transformation connecting the distinct degenerate vacuum states can be constructed from the field operators.  The distinct degenerate vacua belong to distinct orthogonal Hilbert spaces bearing \emph{unitarily inequivalent representations} of the canonical commutation relations.  This is now recognized as a defining feature of spontaneous symmetry breaking.

\section{Electroweak unification}

The 1964 papers from the three groups attracted very little attention at the time.  Talks on the subject were often greeted with scepticism.  By the end of that year, the mechanism was known, and Glashow's (and Salam and Ward's) SU(2) $\X$ U(1) model was known.  But, surprisingly perhaps, it still took three more years for anyone to put the two together.  This may have been in part at least because many of us were still thinking primarily of a gauge theory of strong interactions, not weak.

In early 1967, I did some further work on the detailed application of the mechanism to models with larger symmetries than U(1), in particular on how the symmetry breaking pattern determines the numbers of massive and massless particles \cite{Kibble:1967sv}.  I had some lengthy discussions with Salam on this subject, which I believed helped to renew his interest in the subject.

A unified gauge theory of weak and electromagnetic interactions of leptons was first proposed by Weinberg later that year \cite{Weinberg:1967tq}.  Essentially the same model was presented independently by Salam in lectures he gave at Imperial College in the autumn of 1967 --- he called it the \emph{electroweak theory}.  (I was not present because I was in the United States, but I have had accounts from others who were.)  Salam did not publish his ideas until the following year, when he spoke at a Nobel Symposium \cite{Salam:1968rm}, largely perhaps because his attention was concentrated on the development in its crucial early years of his International Centre for Theoretical Physics in Trieste.

Weinberg and Salam both speculated that their theory was renormalizable, but they could not prove it.  An important step was the working out by Faddeev and Popov of a technique for applying Feynman diagrams to gauge theories \cite{Faddeev:1967fc}.  Renormalizability was finally proved by a young student, Gerard 't Hooft \cite{'tHooft:1971rn}, in 1971, a real \emph{tour de force} using methods developed by his supervisor, Martinus Veltman, especially the computer algebra programme \emph{Schoonship}.

In 1973, the key prediction of the electroweak theory, the existence of the \emph{neutral current} interactions --- those mediated by $Z^0$ --- was confirmed at CERN \cite{Hasert:1973ff}.  That discovery led to the award of Nobel Prizes to Glashow, Salam and Weinberg in 1979.  However Ward was left out, even though he had been a co-author on almost all of Salam's papers on the subject, possibly because of the `rule of three', the restriction that the Prize cannot be split among more than three people.  But 't Hooft and Veltman had to wait until 1999 for their Nobel Prizes.

The next major step was the discovery of the $W$ and $Z$ particles at CERN in 1983 \cite{Arnison:1983rp,Banner:1983jy}, which led to Nobel Prizes for the experimenters, Carlo Rubbia and Simon van der Meer.

Meanwhile, during the 1970s and 1980s there had been separate, parallel development of a gauge theory of strong interactions, \emph{quantum chromodynamics} (QCD).  This is a gauged SU(3) theory, which solves the problem of making the interaction short range by a completely different mechanism, not involving spontaneous symmetry breaking.  So we now have the SU(3) $\X$ SU(2) $\X$ U(1) \emph{standard model}, whose predictions have been verified with increasingly impressive accuracy.

\section{The Higgs boson}

In 1964, or 1967, the existence of a massive scalar boson had been a rather minor and unimportant feature.  The important thing was the \emph{mechanism} for giving masses to gauge bosons and avoiding the appearance of massless Nambu--Goldstone bosons.

But after 1983, the Higgs boson began to assume a key importance as the \emph{only} remaining undiscovered piece of the standard-model jigsaw --- apart that is from the last of the six quarks, the \emph{top quark}.  The standard model worked so well that the Higgs boson, or something else doing the same job, more or less \emph{had} to be present.  Finding the boson was one of the main motivations for building the Large Hadron Collider (LHC) at CERN.  Over a period of more than twenty years, the two great collaborations, ATLAS and CMS, have designed, built and operated their two huge and massively impressive detectors.

As is by now well known, their efforts were rewarded in 2012 by the unambiguous discovery of the Higgs boson by each of the two detectors \cite{Aad:2012tfa,Chatrchyan:2012ufa}.  This led to Nobel Prizes for Englert and Higgs in 2013.  Nambu had earlier in 2008 received a Prize for his role in introducing the idea of spontaneous symmetry breaking into particle physics.

This is the end of a chapter, the completion of the standard model.  But it is certainly not the end of the story.  The model has been amazingly successful; almost all the experimental tests have proved positive.  But the model cannot be the last word.  It is in some ways a mess, with around twenty arbitrary parameters such as mass ratios that we cannot predict.  It is not truly a unified model, because the symmetry group has three factors, each with its own coupling strength, though there have of course been efforts to combine the three pieces into a \emph{grand unified theory} with an even larger symmetry that shows up at energies orders of magnitude above the scale accessible by present-day experiments.  This might require the existence of \emph{supersymmetry}, relating fermions to bosons and vice versa, but disappointingly there is as yet no real empirical evidence for its existence.

Moreover, there are quite a few things that the standard model does not explain.  We do not know what constitutes the dark matter revealed by rotation curves of galaxies, or the dark energy responsible for its recently accelerating expansion.  We do not know why the elementary particles come in three families with very similar structure but wildly differing masses.  Nor do we know why the neutrinos have non-zero, though tiny, masses.  And finally of course, we cannot yet find a place for gravity in the standard model.  Perhaps for that we need string theory or M-theory --- or who knows what?

\ack

I am grateful to the organizers of DICE2014, held at Castiglioncello, 15--19 September 2014 for inviting me to give this talk, and also to the organizers of DISCRETE 2014, at King's College London, 2--6 December 2014, who asked me to repeat it.  

I wish especially to acknowledge the great debt I owe to two people who are sadly no longer with us.  Abdus Salam had a huge influence on the direction of my research.  He was a very supportive and inspiring leader of our group, but sadly died prematurely in 1996.  And earlier this year, I lost my erstwhile collaborator and lifelong close friend Gerry Guralnik, who died of a sudden heart attack just after giving a lecture.


\section*{References}

\begin{thebibliography}{99}

\bibitem{Heisenberg:1932} 
  Heisenberg W 1932  
  \"Uber den Bau der Atomkerne I
  \textit{Z.\ Phys.}\  {\bf 77} 1--11
  
\bibitem{Kemmer:1938}
  Kemmer N 1938  
  The charge-dependence of nuclear forces
  \textit{Proc. Camb. Phil. Soc.} {\bf 34} 354--364
  
\bibitem{Gell-Mann:1961} Gell-Mann M 1961  
  \textit{The Eightfold Way: A theory of strong interaction symmetry}, 
  California Institute of Technology Synchrotron Laboratory Report, CTSL-20, unpublished 

\bibitem{Ne'eman:1961} Ne'eman Y 1961
  Derivation of strong interactions from a gauge invariance
  \textit{Nucl.\ Phys.}\  {\bf 26} 222--229

\bibitem{Yang:1954ek}
  Yang CN and Mills RL 1954
  Conservation of isotopic spin and isotopic gauge invariance
  \textit{Phys.\ Rev.\  } {\bf 96} 191--195

\bibitem{Shaw:1955}
  Shaw R 1955
  \textit{Invariance under general isotopic gauge transformations}  
  Cambridge University PhD thesis, part II, chapter III, unpublished.
  
\bibitem{Feynman:1958}
  Feynman RP and Gell-Mann M 1958
  Theory of the Fermi interaction 
  \textit{Phys.\ Rev.}\  {\bf 109} 193--197
 
\bibitem{Sudarshan:1958vf}
  Sudarshan  ECG and Marshak  RE. 1958
  Chirality invariance and the universal Fermi interaction
  \textit{Phys.\ Rev.}\  {\bf 109} 1860--1862

\bibitem{Schwinger:1957} 
  Schwinger JS  1957 
  A theory of the fundamental interactions
  \textit{Annals Phys.}\  {\bf 2}  407--434

\bibitem{Glashow:1961tr}
  Glashow SL 1961
  Partial symmetries of weak interactions
  \textit{Nucl.\ Phys.\  } {\bf 22} 579--588

\bibitem{Salam:1964} 
  Salam A and Ward JC 1964 
  Electromagnetic and weak interactions
  \textit{Physics Letters} {\bf 13} 168--171

\bibitem{Nambu:1960tm}
  Nambu Y 1960
  Quasi-particles and gauge invariance in the theory of  superconductivity
  \textit{Phys.\ Rev.}\  {\bf 117} 648--663

\bibitem{Anderson:1963}
  Anderson PW 1963
  Plasmons, gauge Invariance, and mass
  \textit{Phys.\ Rev.}\  {\bf 130} 439--442

\bibitem{Nambu:1961tp}
  Nambu Y and Jona-Lasinio G 1961
  Dynamical model of elementary particles based on an analogy with superconductivity. I
  \textit{Phys.\ Rev.}\  {\bf 122} 345--358
  
\bibitem{Goldstone:1961eq}
  Goldstone J  1961
  Field theories with superconductor solutions
  \textit{Nuovo Cim.}\  {\bf 19} 154--164

\bibitem{Goldstone:1962es}
  Goldstone J, Salam A and Weinberg S  1962
  Broken symmetries
  \textit{Phys.\ Rev.}\  {\bf 127} 965--970
 
\bibitem{Guralnik:1964a}
  Guralnik GS 1964
  Photon as a symmetry-breaking solution to field theory. I
  \textit{Phys.\ Rev.}\  {\bf 136} B1404--1416

\bibitem{Guralnik:1964b}
  Guralnik GS 1964
  Photon as a symmetry-breaking solution to field theory. II
  \textit{Phys.\ Rev.}\  {\bf 136} B1417--1422

\bibitem{Englert:1964et}
  Englert F and Brout  R 1964
  Broken symmetry and the mass of gauge vector bosons 
  \textit{Phys.\ Rev.\ Lett.}\  {\bf 13}  321--323
  
\bibitem{Higgs:1964ia}
  Higgs  PW 1964
  Broken symmetries, massless particles and gauge fields 
  \textit{Phys.\ Lett.}\  {\bf 12}  132--133
  
\bibitem{Higgs:1964pj}
  Higgs  PW 1964
  Broken symmetries and the masses of gauge bosons 
  \textit{Phys.\ Rev.\ Lett.}\  {\bf 13}  508--509

\bibitem{Guralnik:1964eu}
  Guralnik  GS, Hagen  CR and Kibble  TWB. 1964
  Global conservation laws and massless particles 
  \textit{Phys.\ Rev.\ Lett.}\  {\bf 13}  585--587

\bibitem{Kibble:1967sv}
  Kibble  TWB 1967
  Symmetry breaking in non-Abelian gauge theories 
  \textit{Phys.\ Rev.}\  {\bf 155}  1554--1561

\bibitem{Weinberg:1967tq}
  Weinberg  S 1967
  A model of leptons 
  \textit{Phys.\ Rev.\ Lett.}\  {\bf 19}  1264--1266
  
\bibitem{Salam:1968rm}
  Salam  A. 1968
  Weak and electromagnetic interactions 
  \textit{Elementary particle theory, Proceedings of the Nobel Symposium 
  held in 1968 at Lerum, Sweden} ed N Svartholm
  (Stockholm: Almqvist \& Wiksell) pp 367--377

\bibitem{Faddeev:1967fc}
  Faddeev LD and Popov VN 1967 
  Feynman diagrams for the Yang--Mills field
  Phys.\ Lett.\ B {\bf 25} (1967) 29.
  
\bibitem{'tHooft:1971rn}
  't~Hooft  G 1971
  Renormalizable Lagrangians for massive Yang-Mills fields 
  Nucl.\ Phys.\ B {\bf 35} (1971) 167--188

\bibitem{Hasert:1973ff}
  Hasert   FJ {\it et al}  [Gargamelle Neutrino Collaboration] 1973 
  Observation of neutrino like interactions without muon or electron in the Gargamelle 
  neutrino experiment 
  \textit{Phys.\ Lett.\ B} {\bf 46}  138--140
  
\bibitem{Arnison:1983rp}
  Arnison  G {\it et al}  [UA1 Collaboration] 1983 
  Experimental observation of isolated large transverse energy electrons with associated 
  missing energy at $s^{1/2} = 540$~GeV 
  \textit{Phys.\ Lett.\ B} {\bf 122} (1983) 103--116

\bibitem{Banner:1983jy}
  Banner  M {\it et al}  [UA2 Collaboration] 1983 
  Observation of single isolated electrons of high transverse momentum in events with 
  missing transverse energy at the CERN anti-p p collider 
  \textit{Phys.\ Lett.\ B} {\bf 122}  476--485
  
\bibitem{Aad:2012tfa}
  Aad G {\it et al}\ [ATLAS Collaboration] 2012
 Observation of a new particle in the search for the standard model Higgs boson with the 
 ATLAS detector at the LHC  \textit{Physics Letters B} {\bf 716}  1--29

\bibitem{Chatrchyan:2012ufa}
  Chatrchyan S {\it et al}\  [CMS Collaboration] 2012 
  Observation of a new boson at a mass of 125 GeV with the CMS experiment at the LHC 
  \textit{Physics Letters B} {\bf 716}  30--61


\end{thebibliography}
\end{document}